\documentclass[11pt,twoside]{article}
\usepackage{asp2010}

\resetcounters

\begin{document}

\title{Sequential Chomospheric Brightening: An Automated Approach to Extracting Physics from Ephemeral Brightening}
\author{Michael~S.~Kirk,$^{1,2,3}$ K.~S.~Balasubramaniam,$^{2,3,1}$ Jason~Jackiewicz,$^{1}$ R.~T.~James McAteer,$^{1}$ Bernie~J.~McNamara$^{1}$
\affil{$^1$ Department of Astronomy, New Mexico State University, P.O. Box 30001, MSC 4500, Las Cruces, New Mexico 88003-8001; mskirk@nmsu.edu}
\affil{$^2$ Space Vehicles Directorate, Air Force Research Laboratory, Kirtland AFB, NM 87114}
\affil{$^3$ National Solar Observatory, Sunspot, NM 88349}}

\begin{abstract}
We make a comparison between small scale chromospheric brightenings and energy release processes through examining the temporal evolution of sequential chromospheric brightenings (SCBs), derive propagation velocities, and propose a connection of the small-scale features to solar flares. Our automated routine detects and distinguishes three separate types of brightening regularly observed in the chromosphere: plage, flare ribbon, and point brightenings. By studying their distinct dynamics, we separate out the flare-associated bright points commonly known as SCBs and identify a propagating Moreton wave. Superimposing our detections on complementary off-band images, we extract a Doppler velocity measurement beneath the point brightening locations. Using these dynamic measurements, we put forward a connection between point brightenings, the erupting flare, and overarching magnetic loops. A destabilization of the pre-flare loop topology by the erupting flare directly leads to the SCBs observed. 
\end{abstract}

\section{Introduction}
\label{S-Intro}
A study of dynamics in the solar chromosphere routinely produces three distinct variants of small scale intensity brightenings (flare-, plage-, and compact-brightenings) all of which are characterized by enhanced H$\alpha$ brightness above the background quiet Sun. The spatial and temporal evolution results from different physical processes responsible for each type of brightening. Typically, brightenings have been identified and characterized manually~\citep[e.g.,][]{2000ASPC..206..426K,1989SoPh..123..309R,2002ESASP.477..187V}.  

One type of flare-associated compact brightenings is called sequential chromospheric brightenings (SCBs). Using high resolution H$\alpha$ images, SCBs were first observed in 2005 and appear as a series of spatially separated points that begin next to the flare, brighten in sequence, and propagate several hundred arc-seconds away from the flare \citep{2005ApJ...630.1160B}. SCBs are most often observed as multiple strings of brightenings in association with a large-scale chromospheric eruptions. The loci of SCBs appear most prevalently along the axis of the flare ribbons and are interpreted as progressive propagating disturbances. \citet{2007AdSpR..39.1781P} demonstrate that SCBs have physical properties consistent with chromospheric evaporation and represent footpoints of field lines that extend into the corona.

\subsection{Data}
\begin{figure} 
	\plotone{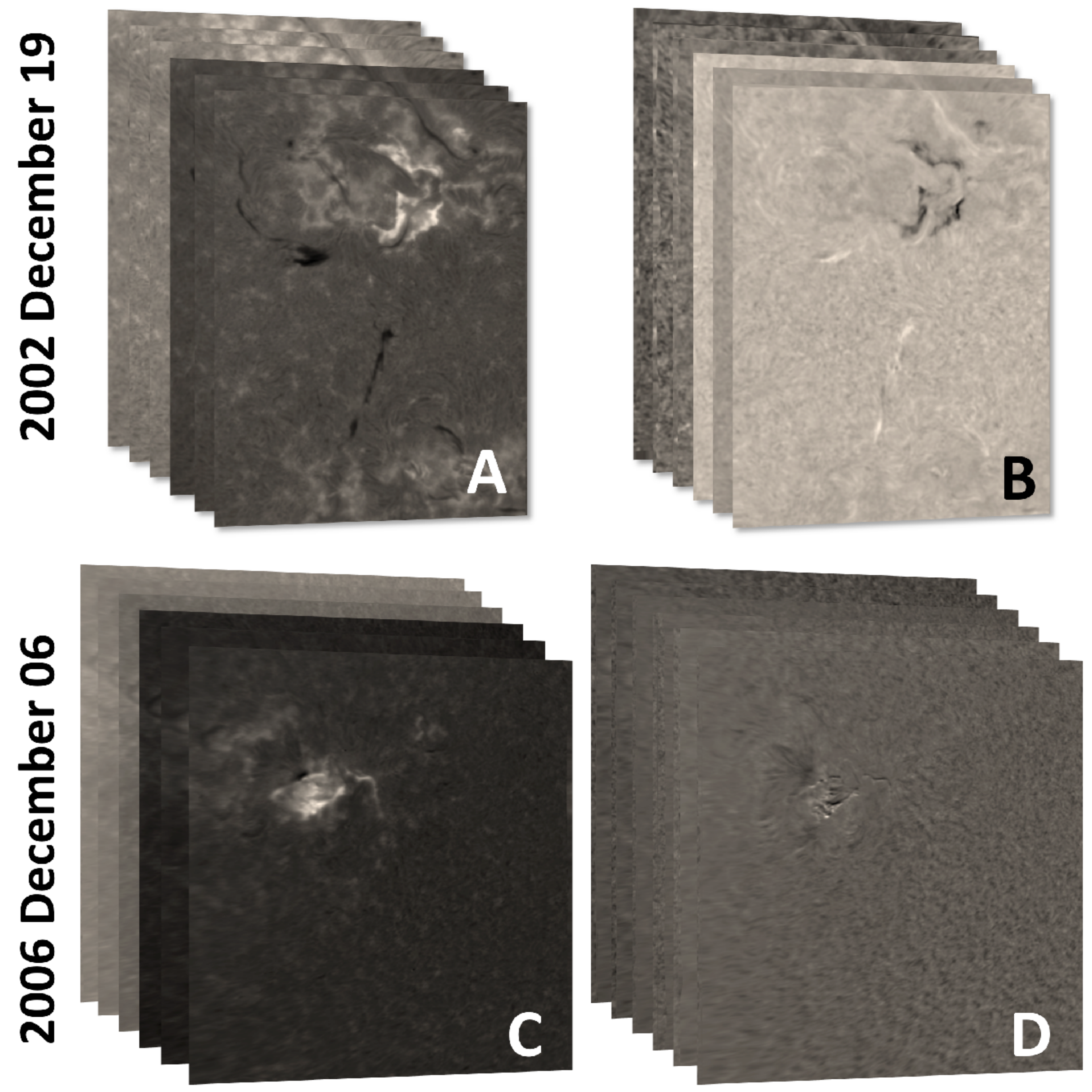}
	
         \caption{The two flaring events in this study: 19 December 2002 and 6 December 2006. {\bf A} shows an example of a region of interest (ROI) time series of calibrated H$\alpha$ ISOON images from 2002. Each image is 633 $\times$ 845 arcsec corresponding to $2.81 \times 10^{11}$ km$^{2}$ on the solar surface. {\bf B} is a Doppler measure of the same ROI as image {\bf A}. The Doppler velocity image is created using a subtraction technique out of ISOON H$\alpha$ off-band images described in Section~\ref{S-Intro}. The velocity ranges from -92.7 to 47.0 km s$^{-1}$ from black to white. {\bf C} shows an example of an ROI time series of calibrated H$\alpha$ ISOON images from 2006. Each image is 990 $\times$ 990 arcsec corresponding to $5.15 \times 10^{11}$ km$^{2}$ on the solar surface. {\bf D} is a Doppler measure of the same ROI as image {\bf C}. The velocity ranges from -84.7 to 802.5 km s$^{-1}$ from black to white. Because {\bf B} and {\bf D} are plotted on different scales, they are not directly comparable.}
   \label{Images}
   \end{figure}

Chromospheric H$\alpha$ (6562.8 \AA) images from the USAF's {\it Improved Solar Observing Optical Network} (ISOON;~\citealp{1998ASPC..140..519N}) prototype telescope are used to study flare ribbons and SCBs. ISOON is an automated telescope producing 2048$\times$2048 pixel full-disk images at a one-minute cadence. Each image has a 1.1 arc-second spatial sampling, is normalized to the quiet Sun, and corrected for atmospheric refraction (Figure~\ref{Images}a and c). At 3 -- 4 seconds after the line center images are taken, ISOON also records H$\alpha$ $\pm0.4$ \AA\ off-band Doppler images. Figure~\ref{Images} (b and d) shows examples of Doppler signals derived from the ISOON red and blue wing images \citep{Kirk2012}. 

For this study, we chose to apply the brightening detection algorithms to two flares with a visually significant number of ephemeral chromospheric brightenings (Table~\ref{T-events}). Both of the events selected has a two-ribbon configuration and an associated halo CME. The 19 December 2002 event is imbedded in a complex active region and the flare ribbons exhibit apparent motion toward a second active region $\approx 9$ arcmin south. The 6 December 2006 event has an associated Moreton wave previously characterized by \citet{2010ApJ...723..587B}. Images covering the entire flaring event are extracted from the archive, yielding a data cube with 250 -- 500 images for each event.  Each image is preprocessed as described in \citet{2011SoPh..tmp..345K} and a region of interest (ROI) is extracted: the 2002 event has an ROI of 633 $\times$ 845 arcsec and the 2006 event an ROI of 990 $\times$ 990 arcsec.

\begin{table}
\caption{ The events selected for study. Each event was visually identified as having a two-ribbon configuration and significant off-ribbon compact brightening associated with the flare eruption. }
\label{T-events}

\begin{tabular}{lccccc}    
  \hline             
 Date & Start Time (UT) & Duration (h) & Flare Class & CME & Wave \\
  \hline
19 Dec 2002& 21:34 & 0.8 & M2.7 & Halo & None Seen\\
6 Dec 2006& 18:29 & 1.5 & X6.5& Halo & Moreton\\
  \hline
\end{tabular}
\end{table}

\subsection{Feature Tracking}
An animated series of images of the ROI reveals several physical characteristics of evolving ribbons: the ribbons separate, brighten, and change their morphology. Adjacent to the eruption, SCBs can be observed brightening and dimming in the region surrounding the ribbons.  In the 2006 event, a Moreton wave can be seen propagating away from the flare eruption.  \citet{2011SoPh..tmp..345K} describe in detail techniques and methods used to identify, track, and extract physical properties such as location, velocity, and intensity of flare ribbons and SCBs. In general, the detection and tracking algorithms are tuned to the ISOON data and specialized for each feature of interest. This specialization requires physical knowledge (e.g. size, peak intensity, and longevity) of that feature being detected to isolate the substructure. The process of identifying and tracking features results in a series of kernels associated with each feature. In this context, a kernel is a small locus of pixels that are associated with each other through increased intensity as compared with the immediately surrounding pixels. This article presents a description of the dynamical properties of flare- and compact-brightenings resulting from applying a new automated method~\citep{2011SoPh..tmp..345K} of identifying and tracking these ephemeral events. 

\section{Small-Scale Flare Dynamics} 
\label{S-Characteristics}
\begin{figure}    
\plotone{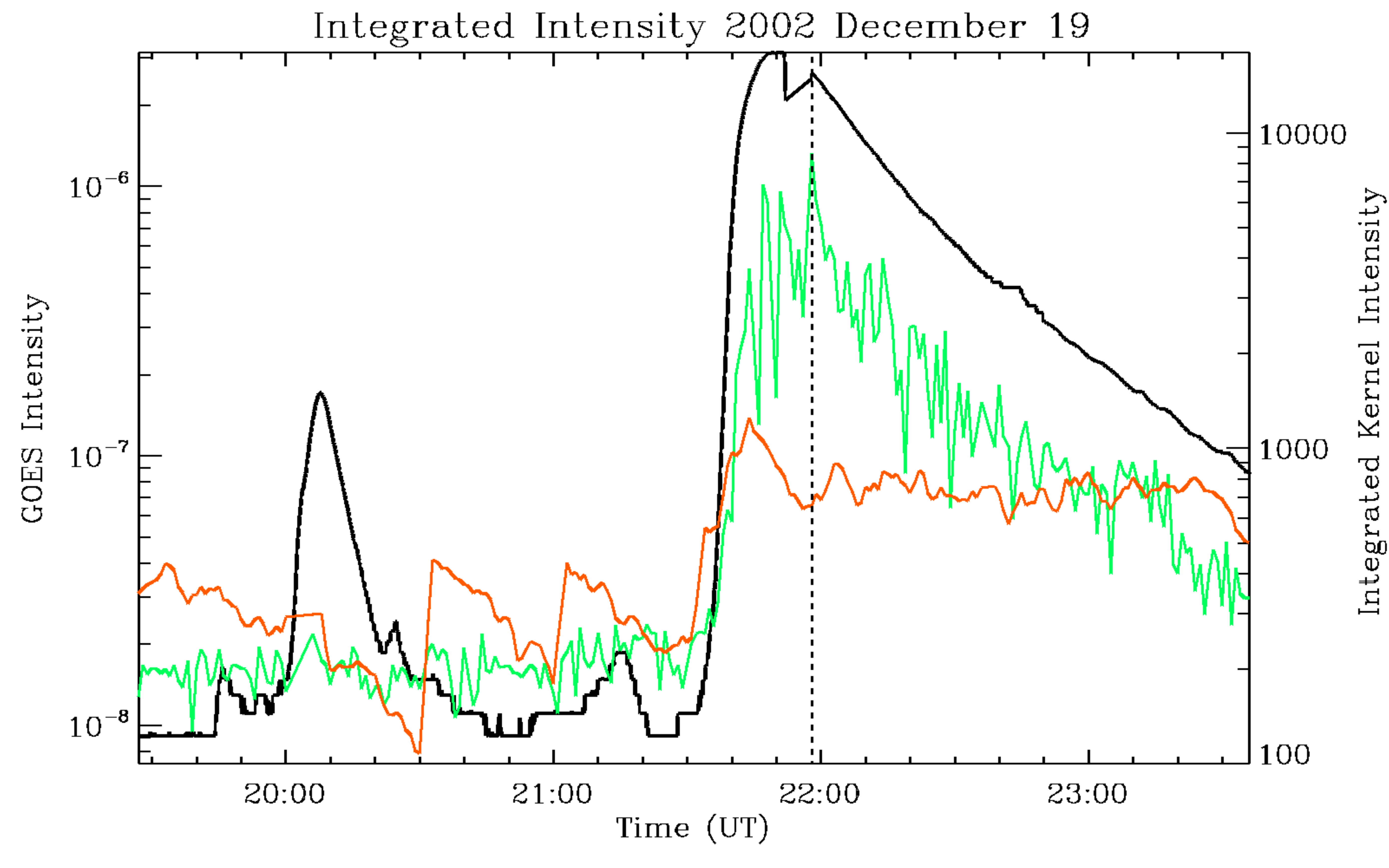}
\plotone{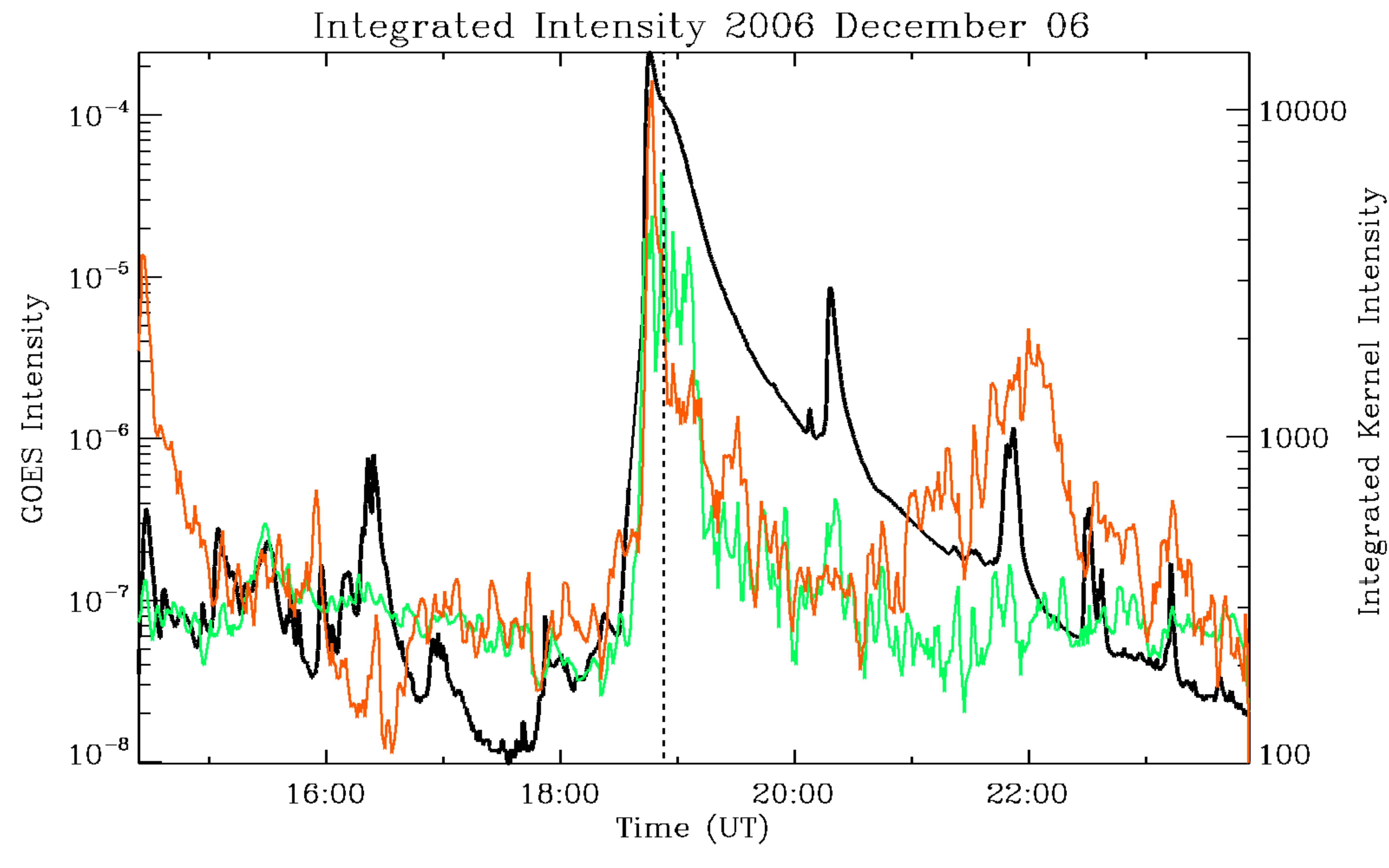}   

             \caption{The 19 December 2002 (top) and 6 December 2006 (bottom) integrated flare curves. The green line is the integrated flare kernel intensity. This curve is generated by summing the intensities of all flare kernels at each time step. The red line is the integrated SCB kernel intensity. This curve is generated by summing the intensities of all SCB kernels at each time step. Overlaid for comparison in black is the GOES $0.5-4.0$ \AA\ intensity curve. The vertical dashed line indicates the peak of the integrated flare curve (green). The integrated flare kernels reproduce the shape of the flare curve determined by the GOES satellite. Also notice that the integrated SCB intensities peak in the impulsive phase of the associated flare.}
   \label{Curves}
   \end{figure}

Integrating the intensity of all the tracked flare kernels at each time step reproduces the overall topology of the flare (Figure~\ref{Curves}) The resultant aggregate intensity curve has the same temporal profile as the GOES 0.5 -- 4.0 \AA\ intensity curve in the impulsive phase, peak, and exponential decline. This indicates that the flare is well characterized by the flare kernels when taken in aggregate, thereby confirming the results of \citet{2011SoPh..tmp..345K}. 

Sequential chromospheric brightenings have a different intensity profile than the flare kernels. Figure~\ref{Curves} also shows the integrated intensities of all SCBs as a function of time for both events studied.  The aggregate SCB intensity curve remains bright for a relatively small time window before decaying to a stable intensity profile. SCBs brighten impulsively about 10 -- 20 minutes before flare peak and decay in intensity after 20 -- 40 minutes. In contrast, the flare intensity curve remains above pre-flare levels for several hours. The integrated SCB curve in the 2002 event peaks in the impulsive phase of the flare but does not return to pre-flare intensity levels for the duration of observation. The 2006 event is significantly more complicated. At the outset of the flare, a Moreton wave is produced. This wave creates a sharp spike in the integrated SCB intensity which dissipates in $\approx 5$ minutes. Around 22:00~UT, in the decay phase of the associated flare, the SCB curve has another spike in intensity. This spike is also captured by the GOES 0.5 -- 4.0 \AA\ intensity curve but is missed by the integrated flare curve indicating another event in the ROI but outside the identified flare ribbons. 

\begin{figure}   
	\plotone{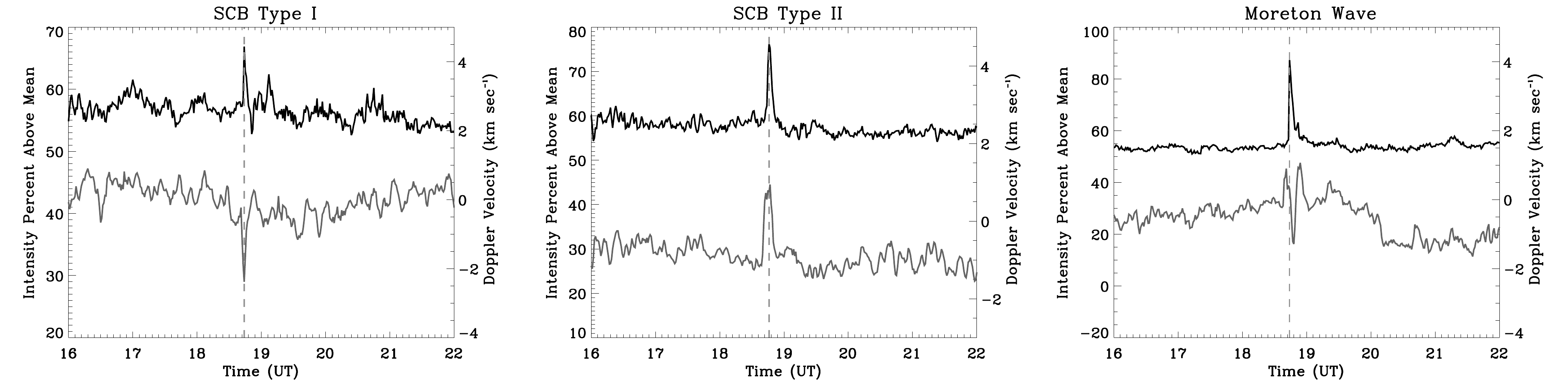}
              \caption{The H$\alpha$ line center intensity (black curve) and Doppler velocity (gray curve) measurements for two different types of SCBs observed in both 2002 and 2006 as well as the Moreton wave observed in the 2006 flare. The vertical dashed line marks the peak H$\alpha$ intensity of the associated compact brightening. A negative Doppler velocity is away from the observer and into the Sun. }
   \label{Types}
   \end{figure}

Examining the Doppler velocity measurements underneath several SCB locations reveals two distinct types of SCBs (Figure~\ref{Types}). A type I SCB has an impulsive intensity profile and an impulsive negative Doppler shift of 2 -- 6 km s$^{-1}$ that occurs close in time to the peak brightening (simultaneously or a few minutes before or after the peak). A negative velocity is defined here as motion away from the observer and into the Sun. A type II SCB has a positive Doppler velocity perturbation of between 2 -- 4 km s$^{-1}$ that often lasts longer than the emission in the H$\alpha$ intensity profile. The timing of the Doppler shift and intensity peak in type II SCBs are nearly coincidental. These types of SCBs are more fully defined by \citet{Kirk2012}. Also shown in Figure~\ref{Types} is the intensity and Doppler profile of the Moreton wave in the 2006 flare. Similar to a type III SCB identified by \citet{Kirk2012}, the Moreton wave demonstrates variable dynamics. It has a impulsive H$\alpha$ intensity line center like other SCBs. However, the Doppler profile of the Moreton wave is distinct. The Moreton wave begins with a positive Doppler shift, then there is an impulsive negative velocity shift followed by a shift back to a positive velocity before it decays to background levels. The magnitude of Doppler velocities associated with the Moreton wave decays as a function of distance from the flare from $\approx 10$ near the flare center down to $\approx 2$ km s$^{-1}$ at the limits of detection.
 
\section{Large-Scale Flare Dynamics}
\label{S-Propagation}

\begin{figure}    
\plotone{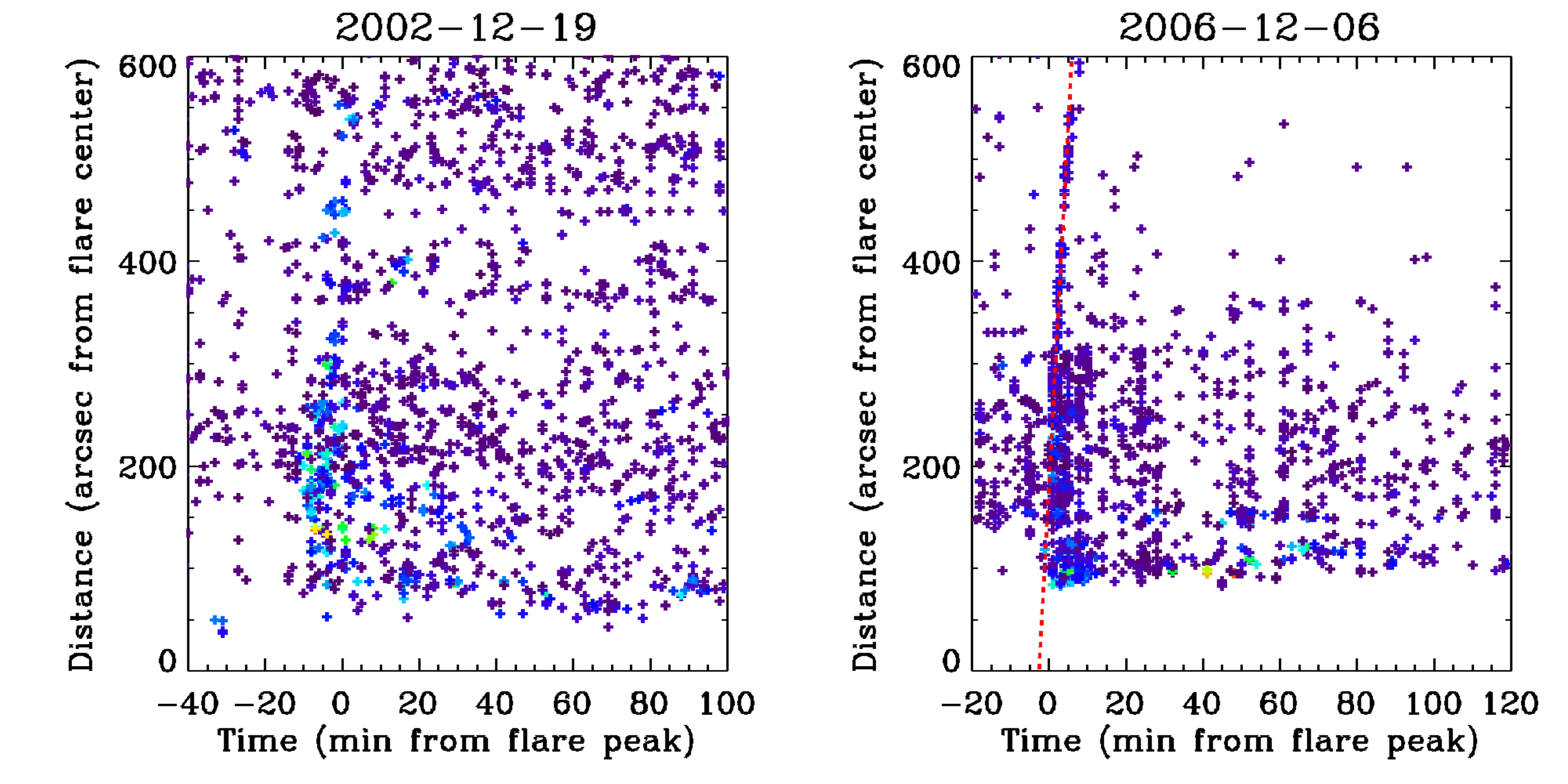}
\plotone{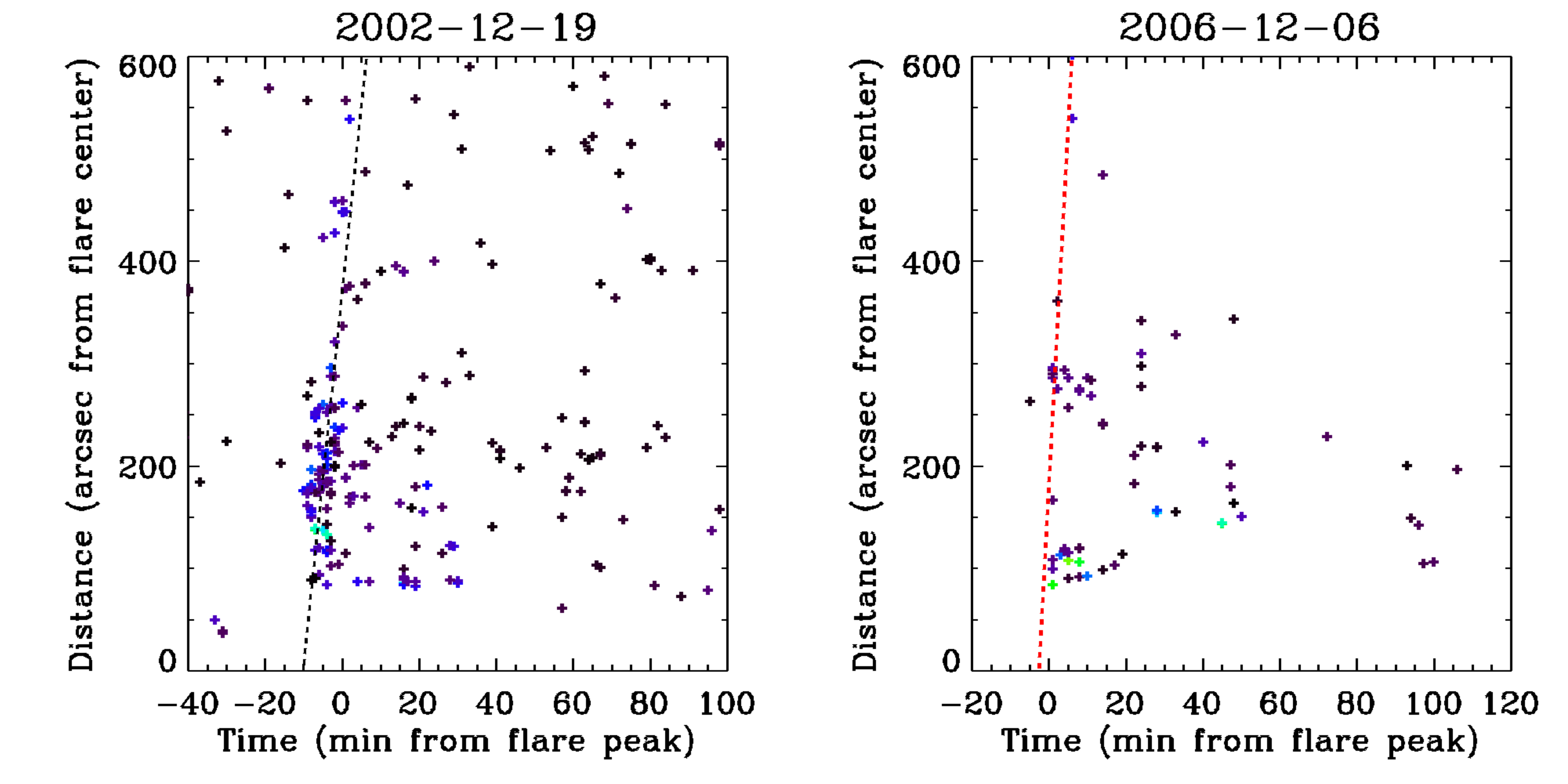}  
              \caption{Top: Derived statistics for all compact brightenings for the 19 December 2002 and 6 December 2006 events. The red dashed line in the 2006 event plots the mean Moreton wave progression derived by \citet{2010ApJ...723..587B}. All panels show the distance at which the bright point occurs versus time from the H$\alpha$ flare peak. The color of each plotted point is representative of its relative intensity; the dimmest are purple while higher intensity detections are yellow. Bottom: the same plots as the top two panels except with a Doppler filter applied -- each point had to have a SCB type I Doppler shift associated with the line center brightening.  The black dashed line on the 2002 event shows a weighted regression fit. In contrast, the red dashed line in the 2006 event again plots the Moreton wave \citet{2010ApJ...723..587B}. }
   \label{Dynamics}
   \end{figure}

By examining the distance of each SCBs from the flare center as a function of time, the propagation trends of SCBs around the flare are exposed (Figure~\ref{Dynamics}). The top two panels of Figure~\ref{Dynamics} show the timing of the SCB peak brightening as a function of its distance from flare center using the technique defined by \citet{Kirk2012}. The first feature that dominates the plots is the sheer number of detections. This is indicative of the complex nature of the active regions studied. Both events are surrounded by interconnected active regions which are strongly influenced by the erupting flare. As the flare erupts, the overall number of compact brightenings increase significantly, especially the number of brightenings within 100 arcsec of the flare. In the 2002 event, the brightest SCBs (lightest symbols) are concentrated around the peak of the flare. In the 2006 event, there is a strong linear feature starting at the flare peak and propagating almost vertically in the plot. This feature tracks the Moreton wave produced in the 2006 eruption.  

Identifying type I SCBs within the total set of compact brightenings tracked eliminates a significant number of detections (Figure~\ref{Dynamics}, bottom panels). A filter is applied to find features with negative Doppler shifts within 3 minutes of the H$\alpha$ peak intensity. With this method of filtering, some of the bright points associated with the Moreton wave in 2006 meet this criteria. The points that are left show fast propagating tenuous trains of brightening in both events.  To illuminate the apparent propagation in the 2002 event, a weighted linear regression technique is used to fit the dashed line shown to the data. The points are weighted by their temporal proximity to flare peak; detections that occurred near flare maximum were given more significance. The resultant fit has a slope of $\approx 440$ km s$^{-1}$ and visually corresponds to an over density of detections in the unfiltered plot above. The fitted line in the filtered plot from the 2006 event is the propagation distance as a function of time of the Moreton wave derived in \citet{2010ApJ...723..587B}. This derived wave propagation has a velocity of $\approx 850$ km s$^{-1}$ and a visual correlation with the linear feature observed in the plot above.

\section{Discussion}
\label{S-Discussion}
The results presented here suggest tracking kernels through the evolution of the erupting flare can characterize several different physical phenomena that make up the evolving active region. We confirm the results of \citet{2011SoPh..tmp..345K} that a sum of the kernels tracked within a flare ribbon reproduces the total intensity curve of the flare. Tracking ephemeral brightenings surrounding the flare aggregates several different physical phenomena. To separate these different events, we use Doppler velocities. This kernel tracking technique is able to follow the progression of a Moreton wave with a sufficient number of detections to carefully characterize its deceleration. 

SCBs are a special case of chromospheric compact brightening that occur in conjunction with flares and CMEs. The distinct nature of SCBs differentiates it from a Moreton wave by unique Doppler velocity profiles and significantly slower propagation speeds. The Moreton wave in 2006 was observed to propagate at nearly twice the speed of the type I SCBs characterized in the 2002 event; an observed speed of $\approx 850$ versus $\approx 440$ km s$^{-1}$.  This confirms the phenomenological difference between SCBs and Moreton waves despite their similar H$\alpha$ intensity profiles. 

The origin of SCBs in the impulsive phase of the flare's evolution, and rapid dispersal, suggests that SCBs are triggered by non-localized phenomena and are indicative of the conditions of the entire flaring region. SCBs can theoretically be caused by a mechanism in which overlying loop lines are destabilized and accelerate electrons to impact a denser chromosphere. In this understanding, SCBs are dynamically triggered by the associated flare but are distinct from the flare arcade itself. In future work, modeling energetic differences between SCBs and flares is needed to explain the coupled phenomena.

\acknowledgments
The authors thank: (1) USAF/AFRL Grant FA9453-11-1-0259, (2) NSO/AURA for the use of their Sunspot, NM facilities, (3) AFRL/RVBXS, (4) Crocker and Weeks for making their algorithm available online, and (5) ATST-EAST for travel support. 

\bibliography{Kirk}

\begin{thebibliography}{}
\expandafter\ifx\csname natexlab\endcsname\relax\def\natexlab#1{#1}\fi
\expandafter\ifx\csname url\endcsname\relax
  \def\url#1{\texttt{#1}}\fi
\expandafter\ifx\csname urlprefix\endcsname\relax\def\urlprefix{URL }\fi
\providecommand{\eprint}[2][]{\url{#2}}

\bibitem[{{Balasubramaniam} et~al.(2010){Balasubramaniam}, {Cliver}, {Pevtsov},
  {Temmer}, {Henry}, {Hudson}, {Imada}, {Ling}, {Moore}, {Muhr}, {Neidig},
  {Petrie}, {Veronig}, {Vr{\v s}nak}, \& {White}}]{2010ApJ...723..587B}
{Balasubramaniam}, K.~S., {Cliver}, E.~W., {Pevtsov}, A., {Temmer}, M.,
  {Henry}, T.~W., {Hudson}, H.~S., {Imada}, S., {Ling}, A.~G., {Moore}, R.~L.,
  {Muhr}, N., {Neidig}, D.~F., {Petrie}, G.~J.~D., {Veronig}, A.~M., {Vr{\v
  s}nak}, B., \& {White}, S.~M. 2010, \apj, 723, 587

\bibitem[{{Balasubramaniam} et~al.(2005){Balasubramaniam}, {Pevtsov}, {Neidig},
  {Cliver}, {Thompson}, {Young}, {Martin}, \&
  {Kiplinger}}]{2005ApJ...630.1160B}
{Balasubramaniam}, K.~S., {Pevtsov}, A.~A., {Neidig}, D.~F., {Cliver}, E.~W.,
  {Thompson}, B.~J., {Young}, C.~A., {Martin}, S.~F., \& {Kiplinger}, A. 2005,
  \apj, 630, 1160

\bibitem[{{Kirk} et~al.(2012){Kirk}, {Balasubramaniam}, {Jackiewicz},
  {McAteer}, \& {Milligan}}]{Kirk2012}
{Kirk}, M.~S., {Balasubramaniam}, K.~S., {Jackiewicz}, J., {McAteer}, R.~T.~J.,
  \& {Milligan}, R.~O. 2012, \apj, Submitted

\bibitem[{{Kirk} et~al.(2011){Kirk}, {Balasubramaniam}, {Jackiewicz},
  {McNamara}, \& {McAteer}}]{2011SoPh..tmp..345K}
{Kirk}, M.~S., {Balasubramaniam}, K.~S., {Jackiewicz}, J., {McNamara}, B.~J.,
  \& {McAteer}, R.~T.~J. 2011, \solphys, 345. \eprint{1108.1384}

\bibitem[{{Kurt} et~al.(2000){Kurt}, {Akimov}, {Hagyard}, \&
  {Hathaway}}]{2000ASPC..206..426K}
{Kurt}, V.~G., {Akimov}, V.~V., {Hagyard}, M.~J., \& {Hathaway}, D.~H. 2000, in
  High Energy Solar Physics Workshop - Anticipating Hess!, edited by {R.~Ramaty
  \& N.~Mandzhavidze}, vol. 206 of Astronomical Society of the Pacific
  Conference Series, 426

\bibitem[{{Neidig} et~al.(1998){Neidig}, {Wiborg}, {Confer}, {Haas}, {Dunn},
  {Balasubramaniam}, {Gullixson}, {Craig}, {Kaufman}, {Hull}, {McGraw},
  {Henry}, {Rentschler}, {Keller}, {Jones}, {Coulter}, {Gregory}, {Schimming},
  \& {Smaga}}]{1998ASPC..140..519N}
{Neidig}, D., {Wiborg}, P., {Confer}, M., {Haas}, B., {Dunn}, R.,
  {Balasubramaniam}, K.~S., {Gullixson}, C., {Craig}, D., {Kaufman}, M.,
  {Hull}, W., {McGraw}, R., {Henry}, T., {Rentschler}, R., {Keller}, C.,
  {Jones}, H., {Coulter}, R., {Gregory}, S., {Schimming}, R., \& {Smaga}, B.
  1998, in Synoptic Solar Physics, edited by {K.~S.~Balasubramaniam, J.~Harvey,
  \& D.~Rabin} (San Francisco), vol. 140 of \aspcs, 519

\bibitem[{{Pevtsov} et~al.(2007){Pevtsov}, {Balasubramaniam}, \&
  {Hock}}]{2007AdSpR..39.1781P}
{Pevtsov}, A.~A., {Balasubramaniam}, K.~S., \& {Hock}, R.~A. 2007, Adv. Space
  Research, 39, 1781

\bibitem[{{Ruzdjak} et~al.(1989){Ruzdjak}, {Vrsnak}, {Brajsa}, \&
  {Schroll}}]{1989SoPh..123..309R}
{Ruzdjak}, V., {Vrsnak}, B., {Brajsa}, R., \& {Schroll}, A. 1989, \solphys,
  123, 309

\bibitem[{{Veronig} et~al.(2002){Veronig}, {Temmer}, {Hanslmeier},
  {Messerotti}, {Otruba}, \& {Moretti}}]{2002ESASP.477..187V}
{Veronig}, A., {Temmer}, M., {Hanslmeier}, A., {Messerotti}, M., {Otruba}, W.,
  \& {Moretti}, P.~F. 2002, in Solspa 2001, Proceedings of the Second Solar
  Cycle and Space Weather Euroconference, edited by {H.~Sawaya-Lacoste}, vol.
  477 of ESA Special Publication, 187

\end{thebibliography}

\end{document}